# Phantom Chain Simulations for Fracture of End-linking Networks

Yuichi Masubuchi

Department of Materials Physics, Nagoya University, Nagoya 4648603, Japan

mas@mp.pse.nagoya-u.ac.jp

ver. Feb. 1, 2024.

**ABSTRACT**

Despite numerous studies, the relationship between network structure and fracture remains unclear. In this study, the fracture properties of end-linking networks were compared with those of loop-free analogs made from star prepolymers by performing phantom chain simulations. The networks were created from equilibrated sols of stoichiometric mixtures of linear prepolymers and $f$-functional linkers through end-linking reactions using Brownian dynamics schemes. The examined networks, with various $f$ values (between 3 and 8) and strand-connection rates ($\varphi_s$), were evaluated in terms of the primary loop fraction and the cycle rank $\xi$. These structural characteristics were consistent with mean-field theories that assume independent reactions. Energy minimization and uniaxial stretch were applied to the networks until they broke without Brownian motion. The fracture characteristics, including strain ($\varepsilon_b$), stress ($\sigma_b$), and work for fracture ($W_b$), were obtained from stress-strain curves. The end-linking networks exhibited larger $\varepsilon_b$ and smaller $\sigma_b$ and $W_b$ than those for star networks due to primary loops, at the same set of $f$ and $\varphi_s$. However, $\varepsilon_b$, $\sigma_b/\nu_{br}$ and $W_b/\nu_{br}$ (with $\nu_{br}$ being the branch point density) lie on the same master curves as those for star networks if they are plotted against $\xi$. This result implies that the fracture of end-linking networks is essentially the same as that for star analogs, and the effects of primary loops are embedded in $\xi$.

**INTRODUCTION**

The effects of structural characteristics of polymer networks on their mechanical properties are a long-standing problem [1–7]. Apart from the effects of strand length distribution, the established concept is that networks become mechanically inferior according to the amount of included structural defects, such as loops, dangling ends, and entanglements[8–12]. Among them, the effects of loops have been widely investigated [8,13–19]. For instance, Zhong et al.[8] proposed the real elastic network theory (RENT) to describe the modulus by the number of loops with elaboratory consideration of higher-order loops. Barney et al.[20] extended the RENT to predict fracture energy by combining the Lake-Thomas theory. These theories have succeeded in describing modulus and fracture energy, which decrease with increasing loops. In particular, primary loops inevitably formed in end-linked networks have a significant impact.[20–23] A class of networks that exclude the formation of primary loops has been made from binary mixtures of star-shaped prepolymers [24–26]. These prepolymers have the same chemistry for the main chains and different end groups that react only between the designated pair. This binary nature eliminates primary loops, and the resultant polymer networks exhibit superior toughness along with the RENT [8].

Despite the success of the RENT [8], an unsolved issue is the effects of node functionality on the fracture. Tsige and Stevens [27,28] performed bead-spring simulations for end-linking networks consisting of linear prepolymers and linkers with various functionalities, and they reported that fracture energy decreases with increasing linker functionality. The RENT [8] could explain this result; the chance of loop formation increases with increasing the linker functionality [17,19,29–32], and such loops depress fracture energy. However, the superiority of small functional networks has also been reported for star-polymer networks without primary loops. Fujiyabu et al. [33] reported that 3-functional networks exhibit superior fracture properties to 4-functional analogous for star polymer networks. Masubuchi et al. [34,35] reported similar results by phantom chain simulations that exclude primary and secondary loops. These studies imply that the number of loops cannot fully explain fracture behavior. It should also be noted that according to the theory by Lin and Zhao [36], fracture energy increases with increasing secondary loops (cyclic loops in their terminology), contradicting the RENT [8].

Apart from the number of loops, another possible approach to characterize network structures is the cycle rank [16,37–39], which describes the modulus in the phantom network theory [39]. Masubuchi et al. [35,40] systematically investigated fracture properties of star polymer networks with various node functionalities $f$ and conversion ratios $\varphi_c$ with the same arm length. They demonstrated that stress and strain at break and work for fracture, $\varepsilon_b$, $\sigma_b$, and $W_b$, and the broken strand fraction $\varphi_{bb}$, lie on master curves if they are summarized as functions of the cycle rank per branch point $\xi$. Masubuchi [41] further showed that the data for star polymer mixtures with different functionalities also obey the same master curves. These results suggest that the cycle rank is an essential parameter for fracture properties. However, the results up to the present are only for star polymer networks, in which primary and odd-ordered loop formations are eliminated, though secondary and even-ordered loops are included. Since the effect of primary loops is known to impact mechanical properties significantly, networks with primary and odd-ordered loops should be examined and compared with the star polymer networks.

In this study, fracture properties of end-linking networks, including loops, were examined via phantom chain simulations, which have been used for star polymer networks. It was found that the obtained fracture properties follow the same relation as those for star polymer networks against $\xi$, implying that the effects of loops can be embedded in $\xi$. Details are shown below.

**MODEL AND SIMULATIONS**

The simulation method is the same as what was used in the previous studies [34,35,40,41], except for considering linear prepolymers and multi-functional linkers instead of star prepolymers. Namely, linear bead spring chains without excluded volume were dispersed in a simulation box and mixed with multi-functional linkers. From such sols, networks were created via end-linking reactions with a Brownian dynamics scheme [42,43]. The resultant networks at various conversion ratios were stretched until the break with an energy minimization scheme without Brownian motion. The fracture characteristics were extracted from the stress-strain relation. Details of the examined systems and the employed simulation schemes are explained below.

The examined systems contained bead spring linear chains with the bead number $N_l$ and cross-linkers with the functionality $f$. They were dispersed in a cubic simulation box with periodic boundary conditions. The numbers of cross-linkers and prepolymers are $M_c$ and $M_p$. Because the stoichiometric condition was considered, $M_p = fM_c/2$. The total bead number of a single cross-linker is $f + 1$, including one central bead and $f$ arm beads. Thus, the total bead number in the system is $M_c(f + 1) + N_l f M_c/2$. The volume of the simulation box was determined to attain the chosen bead density $\rho$. For bead connectivity, a non-linear spring was employed with the spring constant written as $f_{ik} = (1 - \mathbf{b}_{ik}^2/b_{\max}^2)^{-1}$, where $\mathbf{b}_{ik}$ is the bond vector between bead $i$ and $k$, and $b_{\max}$ is the parameter describing the maximum bond length. This finite extensibility was introduced to ensure that bond breakage occurs only due to the elongation process explained later. The sols containing linear prepolymers and linkers were equilibrated by a Brownian dynamics scheme, in which the bead position $\mathbf{R}_i$ obeys the following Langevin equation.

$$\mathbf{0} = -\zeta \dot{\mathbf{R}}_i + \frac{3k_B T}{a^2} \sum_k f_{ik} \mathbf{b}_{ik} + \mathbf{F}_i \qquad (1)$$

Here, $\zeta$ is the friction coefficient, $a$ is the average bond length, and $\mathbf{F}_i$ is Gaussian random force satisfying $\langle \mathbf{F}_i \rangle = \mathbf{0}$ and $\langle \mathbf{F}_i(t)\mathbf{F}_j(t') \rangle = 2k_B T \delta_{ij} \delta(t - t')\mathbf{I}/\zeta$. Hereafter, units of length, energy, and time are chosen as $a$, $k_B T$, and $\tau = \zeta a^2/k_B T$, and quantities are normalized according to these units. After equilibration, with the Brownian motion, end-linking reactions were introduced between chain ends and linker arms when reactive beads came closer to each other than a certain critical distance $r_c$ with a reaction rate $p$. During this gelation process, snapshots of the system were stored at several conversion rates $\varphi_c$.

The obtained networks were subjected to energy minimization with the Broyden-Fletcher-Goldfarb-Sanno method [44]. The minimized potential energy is consistent with the spring force in eq 1 and is written as follows.

$$U = -\frac{3k_B T b_{\max}^2}{2a^2} \sum_{i,k} \ln\left(1 - \frac{\mathbf{b}_{ik}^2}{b_{\max}^2}\right) \qquad (2)$$

To the energy-minimized structure, a stepwise infinitesimal uniaxial affine elongation followed by energy minimization was applied. After each step, the bond length $|\mathbf{b}_{ik}|$ was examined, and when $|\mathbf{b}_{ik}|$ exceeded a critical value $b_c$, the examined bond was removed. Note that $b_c$ is chosen as larger than $b_{\max}$ to avoid bond breakage before elongation. The energy minimization process was repeatedly applied unless no bond breakage occurred without elongation. These processes, including elongation, energy minimization, and bond breakage, were sequentially repeated until network percolation was eliminated.

It is fair to note that the employed simulation scheme cannot consider structural relaxation induced by bond breakage. Simulations with Brownian motion in the elongation and bond breakage processes can deal with such relaxation [43,45]. However, as reported earlier [34,43], such dynamic schemes require several parameters that affect fracture behavior, and the determination of an optimum parameter set, which reproduces experimental conditions, is challenging. This study follows the earlier ones [46–49], eliminating Brownian motion during the elongation process and suppressing the number of parameters.

The simulation parameters were chosen as follows for a direct comparison to the previous simulation study for star polymer networks [35]. The cross-linker functionality $f$ was varied from 3 to 8. The prepolymer bead number $N_l$ was fixed at 8. This $N_l$ and connected linker beads at both sides realize the same strand bead number 10 with the previous star polymer case, where the arm length of star prepolymers was 5. The number of cross-linkers $M_c$ was fixed at 1600. The rest of the parameters were the same as those in the previous study. The bead number density $\rho$ was 8. $b_{\max}$ and $b_c$ were chosen at $\sqrt{2}$ and $\sqrt{1.5}$. The numerical step size for the second-order Brownian dynamics scheme [50] was $\Delta t = 0.01$. The parameters for gelation were set as $r_c = 0.5$ and $p = 0.1$. Concerning the energy minimization process, the energy conversion and bead displacement parameters were $\Delta u = 10^{-4}$ and $\Delta r = 10^{-2}$. For each condition, eight independent simulation runs started from different initial configurations were conducted for statistics.

## RESULTS AND DISCUSSION

Figure 1 shows a series of snapshots of one of the examined systems with $f = 4$. Panel (a) illustrates one of prepolymers (left) and a linker (right). Panel (b) exhibits the gelated network after energy minimization. The conversion ratio $\varphi_c$ for this specific case is 0.975. Panels (c) and (d) show the structure during elongation, and panel (e) displays the structure that loses percolation along the elongational direction after sufficient bond breakage. Panels (f) and (g) exhibit the development of broken strand fraction $\varphi_{bb}$ and true stress $\sigma$ as functions of true (Hencky) strain $\varepsilon$. No essential difference was found in these behaviors from the star polymer analogs reported previously [34,35,40,41].

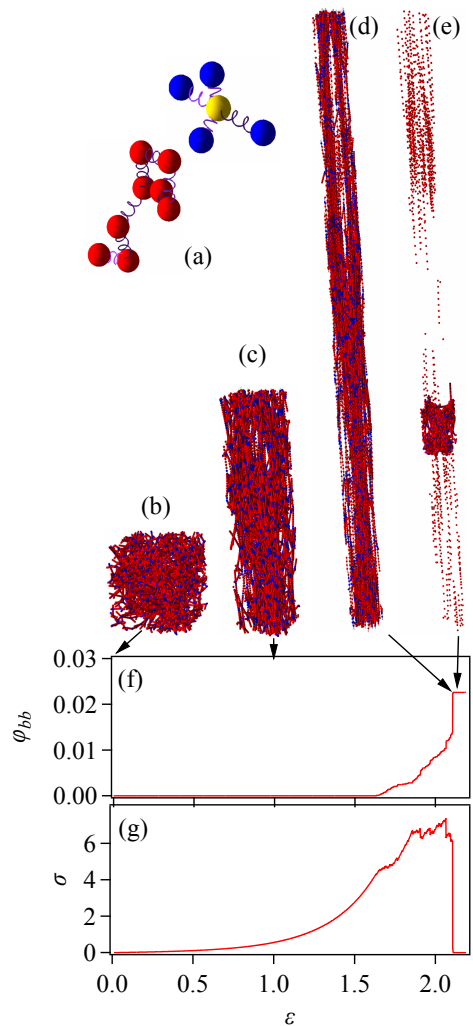

**Figure 1** Snapshots of an examined system with $f = 4$ for the prepolymer and linker (a), the network after gelation with $\varphi_c = 0.975$ and energy minimization (b), the structures under stretch at $\varepsilon = 1.0$ (c) and 2.1 (d), and the broken system at $\varepsilon = 2.11$ (e). In panel (e), dispersed

dots are isolated fragments simultaneously separated from the network at the breakage. Panels (f) and (g) are the development of the broken chain fraction $\varphi_{bb}$ and true stress $\sigma$.

Before examining the fracture properties, let us evaluate the gelated networks before break. Figure 2 (a) shows the fraction of prepolymers for which both ends are reacted, plotted against the conversion ratio $\varphi_c$ for the gelated networks. Namely, the number ratios of free prepolymers and dangling ends, $\varphi_{fc}$ and $\varphi_{dc}$, are subtracted from unity and shown against $\varphi_c$. Because the comparison between end-linking and star polymer networks is one of the motivations of this study, the conversion ratio $\varphi_c$ should be considered accordingly. In the case of star polymer networks, a single reaction connects a pair of branch points. Thus, $\varphi_c$ is identical to the ratio of strand formation, $\varphi_s$. In contrast, for the examined end-linking case, two reactions at both sides of a single linear prepolymer are necessary to connect a pair of branch points. Naively, a relation $\varphi_s = \varphi_c^2 = 1 - (\varphi_{fc} + \varphi_{dc})$ is expected, if loops are included in $\varphi_s$. The broken red curve in Fig 2 (a) shows this relation, and the simulation results are consistent. (Here, all the examined cases for $f = 3 - 8$ overlap.) Other expected relations written as $\varphi_{fc} = (1 - \varphi_c)^2$ and $\varphi_{dc} = 2\varphi_c(1 - \varphi_c)$, were also confirmed. However, the data are not shown because these prepolymers do not contribute to the mechanical response. Hereafter, $\varphi_s$ is used instead of $\varphi_c$ to represent the magnitude of the gelation.

Figure 2 (b) shows the fraction of prepolymers forming primary loops $\varphi_{l1}$ plotted against $\varphi_s$. As expected, $\varphi_{l1}$ increases with increasing $f$ and $\varphi_s$. Naively, $\varphi_{l1}$ can be written as follows.
$$\varphi_{l1} = \varphi_c^2 \{p_{ob}(1 - \varphi_c^{f-1})\}/\{p_{db}(1 - \varphi_c^f)\}, \tag{3}$$
where $p_{ob}$ and $p_{db}$ are probabilities of finding the original branch point and different branch points at the position of the subjected reactive dangling end. The factor $\varphi_c^2$ is the probability of finding a prepolymer with two reacted chain ends. $(1 - \varphi_c^{f-1})$ and $(1 - \varphi_c^f)$ stand for the probabilities of finding at least one vacant reactive site at the examined branch points. (At the original branch point, the examined dangling arm occupies one of the reactive sites; thus, the power is $f - 1$.) Since the examined phantom chain follows the Gaussian statistics, $p_{ob}$ can be found in the literature [31]. $p_{db}$ can be approximately estimated from the average number density of branch points if the density is sufficiently high and structural correlations are neglected. Broken curves in Fig 2 (b) exhibit the results of this calculation being fairly consistent with the simulation results. Figure 2 (c) shows the fraction of prepolymers involved in secondary loops $\varphi_{l2}$. As stated earlier [8], $\varphi_{l1}$ and $\varphi_{l2}$ have one-to-one correspondence, as shown in Fig 2 (d).

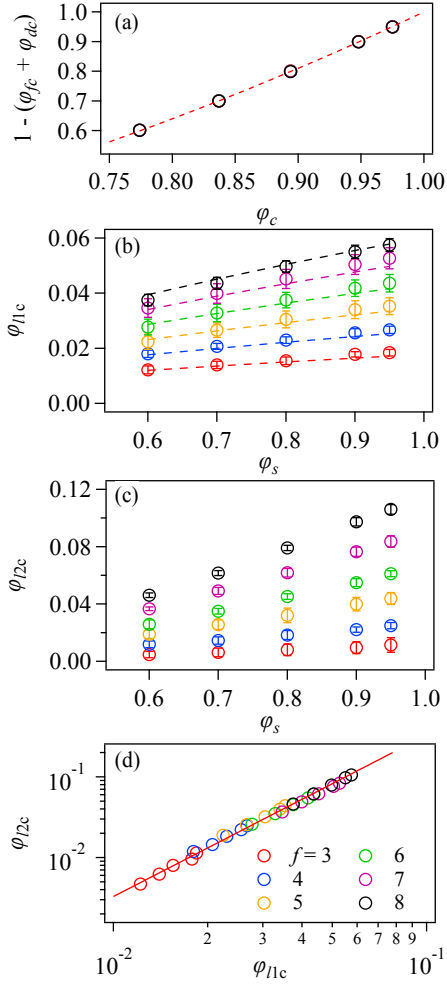

**Figure 2** The ratio of prepolymers with two ends reacted, $1 - (\varphi_{fc} + \varphi_{dc})$, plotted against the conversion ratio $\varphi_c$ (a), the ratio of prepolymers forming primary and secondary loops $\varphi_{l1}$ and $\varphi_{l2}$ plotted against the strand formation ratio $\varphi_s$ (b) and (c), and the relation between $\varphi_{l1}$ and $\varphi_{l2}$ (d). In panel (a), the data for various $f$ overlap and the broken curve shows the relation $1 - (\varphi_{fc} + \varphi_{dc}) = \varphi_c^2$. The cures in Panel (b) show the estimation according to eq 3 for $f$ varied from 3 to 8 from bottom to top. Error bars in Panels (b) and (c) correspond to the standard deviations among eight different simulation runs. The solid line in Panel (d) indicates an apparent relation $\varphi_{l2} = 33\varphi_{l1}^2$.

Figure 3 (a) exhibits the cycle rank per branch point $\xi$ plotted against $\varphi_s$ for various $f$. Due to primary loops, $\xi$ is smaller than the star polymer networks for the same set of $f$ and $\varphi_s$. Thus, for the evaluation of $\xi$, the original mean-field theory [51,52] should be modified by taking account of the primary loop formation. In the original theory (neglecting the loop formation), the probability of finding a strand that is not involved in the effective network, $p_{out}$, is written as

$$p_{out}(f, \varphi_s) = \varphi_s p_{out}^{f-1} + (1 - \varphi_s). \qquad (4)$$

With $p_{out}$, the probability of finding a branch point with $g$ effective arms $p_{eff}(f,\varphi_s,g)$ is written as

$$p_{eff}(f,\varphi_s,g) = {}_fC_g(1-p_{out})^g p_{out}^{f-g}. \quad (5)$$

The number of effective nodes and strands per branch point $\nu$ and $\mu$, and the cycle rank $\xi$ are then given by

$$\nu(f,\varphi_s) = \sum_{i=3}^{f} i p_{eff}(f,\varphi_s,i) \quad (6)$$

$$\mu(f,\varphi_s) = \sum_{i=3}^{f} p_{eff}(f,\varphi_s,i) \quad (7)$$

$$\xi = \nu - \mu \quad (8)$$

This mean-field theory successfully describes $\xi$ for the star polymer networks [35,41]. A straightforward modification of this theory for the examined end-linking cases is using $p_{out}' = p_{out} + \varphi_{l1}$ instead of $p_{out}$. Broken curves in Fig 3 show the results of calculations with $p_{out}'$, demonstrating that the simulation results are excellently described by the modified mean-field theory. Note that the actual $\varphi_{l1}$ is used rather than the estimation shown by eq 3. Nevertheless, the coincidence between the data and the theoretical prediction demonstrates that the examined networks are statistically valid.

Figure 3 (b) shows the Mooney modulus $G$ obtained from the stress-strain curves plotted against $\xi$. Here, $G$ is normalized concerning the branch point density $\nu_{br}$. Since the branch point position changes according to the force balance under stretch, $G$ is expected to follow the phantom network theory [39] and be dominated by the cycle rank. The simulation results are consistent with this idea, although weak nonlinearity is observed for large-$\xi$ due to stretch in energy-minimized networks [34,35]. The effect of primary loops is mostly absorved in $\xi$, but still $G/\nu_{br}$ is slightly smaller than that for star networks (cross) for large-$\xi$.

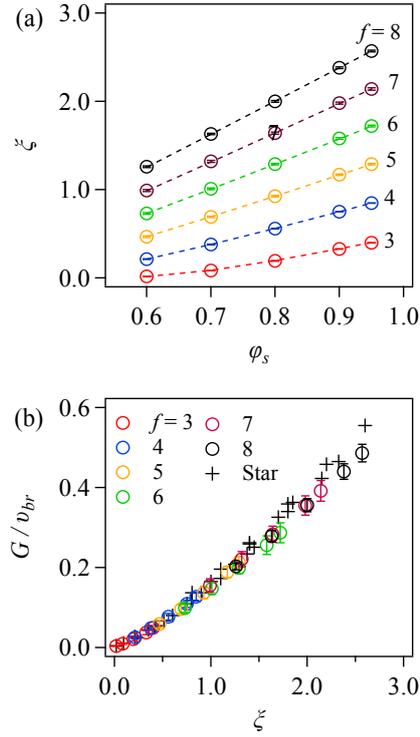

**Figure 3** The cycle rank per branch point $\xi$ plotted against $\varphi_s$ for various $f$ (a) and the Mooney modulus $G$ normalized by the branch point density $\nu_{br}$ plotted against $\xi$ (b). Circles are the simulation results, for which the standard deviations among 8 independent simulation runs are within the symbol size. Broken curves in Panel (a) are the mean-field theory with $p_{out}' = p_{out} + \varphi_{l1}$. The cross in Panel (b) indicates the results of star networks. Error bars exhibit the standard deviations among eight different simulation runs.

Let us turn our attention to the fracture properties. Figure 4 shows stress-strain relationships for the networks with $f = 3$ and $\varphi_s = 0.6$ and $0.95$ (red solid and broken curves) compared to analogs for $f = 8$ (black and gray curves). The network with $f = 8$ is mechanically superior to the one with $f = 3$ when $\varphi_s = 0.6$. However, $f = 3$ becomes better than $f = 8$ for $\varphi_s = 0.95$. This superiority of small $f$ networks at large $\varphi_s$ has also been observed for star polymer networks [35].

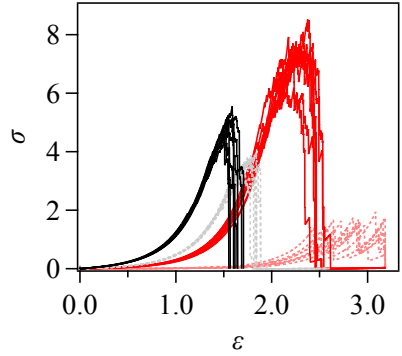

**Figure 4** Stress-strain relations during stretch for $f = 3$ (red curves) and 8 (black and gray curves) at $\varphi_s = 0.6$ (broken curves) and 0.95 (solid curves). Each curve corresponds to different simulation run. $\varepsilon$ is true strain (Hencky strain) and $\sigma$ is true stress.

From the stress-strain behavior, fracture characteristics $\varepsilon_b$, $\sigma_b$, and $W_b$ were collected. ($W_b$ was obtained by numerically integrating the stress-strain curve until the break.) Figures 5 and 6 exhibit these values plotted against $\varphi_s$ and $f$. In Fig 5 (a), $\varepsilon_b$ decreases with increasing $\varphi_s$, and $f = 3$ networks are stretchable compared to $f = 8$ irrespective of $\varphi_s$. In Figs 5 (b) and (c), as mentioned in Fig 4, $f = 8$ is better than $f = 3$ at small $\varphi_s$, but the relation becomes opposite at large $\varphi_s$ with a crossover located between $\varphi_s = 0.7$ and 0.8 for $\sigma_b$ and around $\varphi_s = 0.7$ for $W_b$. In Fig 6 (a), $\varepsilon_b$ decreases with increasing $f$, and as mentioned in Fig 5 (a), smaller $\varphi_s$ networks exhibit larger $\varepsilon_b$. Concerning $\sigma_b$ and $W_b$, these values monotonically decrease with increasing $f$ for $\varphi_s = 0.9$. In contrast, they show a peak at a certain $f$ when $\varphi_s$ is small, and a decay follows. Due to the inclusion of primary loops, $\varepsilon_b$, $\sigma_b$, and $W_b$ differ from those for the star polymer analogs, for which triangles show the data [35]. Namely, $\varepsilon_b$ is larger, whereas $\sigma_b$, and $W_b$ are smaller than stars. However, the difference is not significant, and the effects of $\varphi_s$ and $f$ on the network fracture are essentially the same as what was found for the star polymer networks; large $f$ networks are robust against a change of $\varphi_s$, whereas small $f$ networks exhibit superior toughness if $\varphi_s$ becomes sufficiently large.

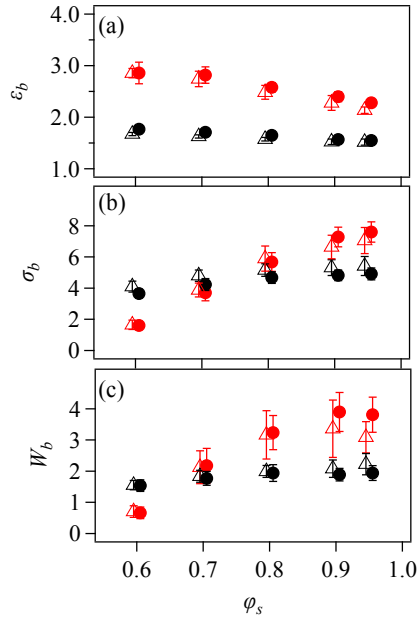

**Figure 5** Fracture characteristics $\varepsilon_b$, $\sigma_b$, and $W_b$ from top (panel a) to bottom (panel c) plotted against $\varphi_s$. Red and black symbols are for the results with $f = 3$ and 8, respectively. Circles and triangls shows the results for end-linking and star networks, respectively. Error bars indicate standard deviations among eight different simulation runs. The data for star networks are reported in the previous study [35].

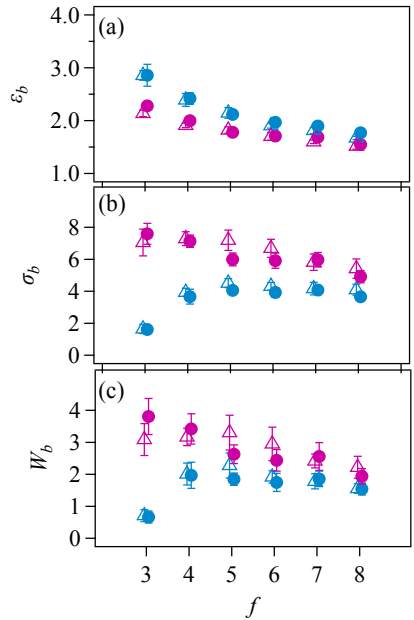

**Figure 6** Fracture characteristics $\varepsilon_b$, $\sigma_b$, and $W_b$ from top (panel a) to bottom (panel c) plotted against $f$. Skybule and violet symbols are for the results with $\varphi_s = 0.6$ and 0.95, respectively.

Circles and triangles shows the results for end-linking and star networks, respectively. Error bars indicate standard deviations among eight different simulation runs. The data for star networks are reported in the previous study [35].

Apart from the abovementioned qualitative comparison between end-linking and star networks, an interesting argument is whether these networks have fundamental differences in the relation between structural and fracture characteristics. Masubuchi et al.[35,41] reported fracture characteristics for star networks with various $f$ and $\varphi_s$ can be summarized on master curves if they are plotted against $\xi$. Let us see if the results of end-linking networks exhibit differences in such an analysis.

Figure 7 shows $\varepsilon_b$, $\sigma_b$, and $W_b$ plotted against $\xi$ for all the obtained results with various $f$ and $\varphi_s$. Panel (a) demonstrates that $\varepsilon_b(\xi)$ follows the same master curve as that exhibited for star networks, for which the black crosses show the results. Thus, the difference in $\varepsilon_b$ from star polymers seen in Figs 5 and 6 can be explained by the difference of $\xi$. Panels (b) and (c) show $\sigma_b$ and $W_b$ normalized by the number density of branch points $\nu_{br}$. Note that $\nu_{br}$ is calculated from all the existing linkers in the system; even unreacted ones are included. In the previous studies [35,40,41], $\sigma_b$ and $W_b$ were normalized by the broken bond fraction $\varphi_{bb}$. However, $\varphi_{bb}$ is hardly accessible experimentally, and normalization according to $\nu_{br}$ is proposed here instead, confirming that the results for star networks lie on master curves, as demonstrated by the black crosses. The end-linking network results shown by circles follow the same master curves as star networks, revealing that the difference from star networks is due to the difference in $\xi$, as demonstrated for $\varepsilon_b$. One may argue if the master curves for $\sigma_b/\varphi_{bb}$ and $W_b/\varphi_{bb}$ are the same as those for star networks. Panel (d) shows that $\varphi_{bb}/\nu_{br}$ for end-linking and star networks draw the same master curve, implying that $\sigma_b/\varphi_{bb}$ and $W_b/\varphi_{bb}$ for the end-linking case also lie on the same master curve as those for star networks, although data are not shown here.

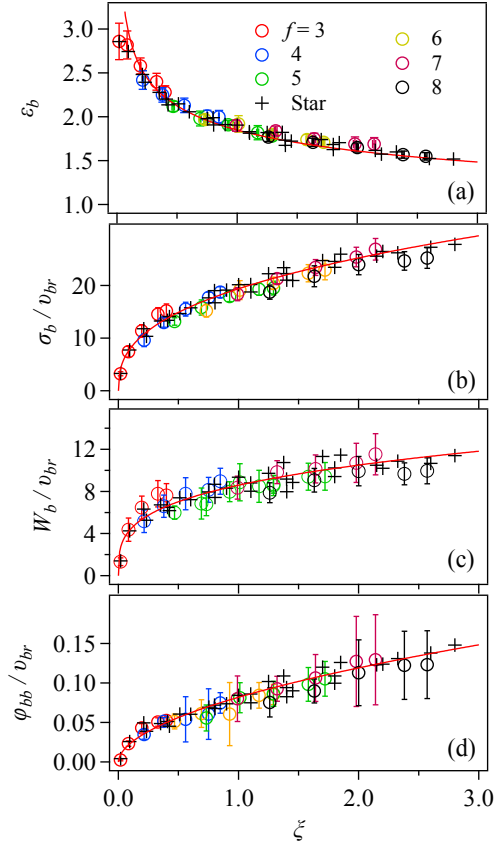

**Figure 7** Fracture characteristics plotted against $\xi$. Solid curves indicate apparent relations written as $\varepsilon_b = 1.85\xi^{-0.18}$, $\sigma_b/\nu_{br} = 19.4\xi^{0.34}$, $W_b/\nu_{br} = 8.6\xi^{0.29}$, and $\varphi_{bb}/\nu_{br} = 0.082\xi^{0.54}$, from panel (a) to (d). Error bars display standard deviations among 8 different simulation runs. The data for star networks shown by cross are reported in the previous study [35].

The physics behind the observed universality for fracture characteristics as functions of cycle rank is yet to be clarified. (Thus, it is fair to note that the power-law relations shown in Figure 7 are purely empirical eye-guides, and other functional forms are also applicable.) A possible qualitative explanation attempted in the previous studies [35,41] is related to strand-extending prepolymers that are connected to linkers with only two effective connections. Such prepolymers contribute to extending the strand length between a pair of effective network nodes and enlarging the involved closed cycle. A question is how the primary loop formation affects such extending chains. Concerning such an argument, Fig. 8 exhibits the number-averaged molecular weights of effective and broken strands, $M_n$ and $M_{bn}$, as functions of $\xi$ normalized by the strand molecular weight for the case with a single prepolymer connecting two branch points, $M_0$. Here, $M_{bn}$ means the value of $M_n$ before the breakage averaged only for the broken strands at the breakage. For the case of star polymer networks (shown by cross), apparent relations have been reported written as $M_n/M_0 = 1 + e^{-4\xi}$ and $M_{bn}/M_0 = 1 + 0.5e^{-4\xi}$, describing that $M_{bn} < M_n$. As explained previously, this result demonstrates that strand-extending chains are relatively unbroken due to their large-$M_n$. Concerning the end-linking case (indicated by circles), $M_n/M_0$ is smaller than star networks for small $f$ cases. See panel (a). This small-$M_n/M_0$ is due to primary loop formation. As seen in Fig 3, primary loops suppress $\xi$ of end-linking networks compared to that for star analogs at the same set of $f$ and $\varphi_s$. Thus, if the networks are compared for a certain $\xi$, the end-linking one should have larger $f$ and $\varphi_s$ that realize small $M_n/M_0$. In contrast, $M_{bn}/M_0$ in panel (b) is larger than that for star networks and even larger than $M_n/M_0$ when $f$ is larger than 4. In particular, the case with $f = 4$ (blue circle) exhibits $M_{bn} > M_n$. This result demonstrates that extended strands have a bias for breakage contrary to those in the star analogs. Nevertheless, these qualitative differences in $M_n/M_0$ and $M_{bn}/M_0$ do not appear in the fracture characteristics as functions of $\xi$ shown in Fig 7, and the universality in the fracture characteristics seem unrelated to the strand-extending chains.

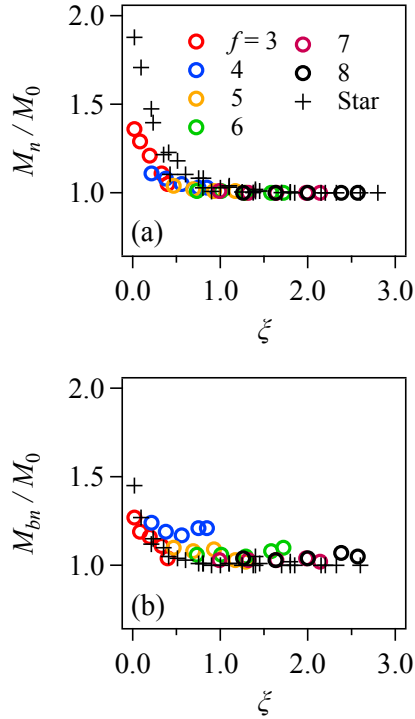

**Figure 8** Number averaged effective strand molecular weight $M_n$ normalized by the molecular weight of the single strand $M_0$ for the entire effective strands (a) and that for the broken strands $M_{bn}$ (b). The data for star networks shown by cross are reported in the previous studies [35,41].

**CONCLUSIONS**

Phantom chain simulations were performed for the fracture of end-linking networks. Sols of linear prepolymers and $f$-functional linkers were equilibrated and gelated with various strand-connection rates $\varphi_s$ through Brownian dynamics schemes. The created networks were statistically valid regarding the primary loop fraction and the cycle rank $\xi$, in the sense that these values were consistent with mean-field calculations in which all the reactions occur independently. Energy minimization and uniaxial stretch were applied to such networks until the break. From the stress-strain relation observed during the stretching process, fracture characteristics, such as strain and stress at break $\varepsilon_b$ and $\sigma_b$ and work for fracture $W_b$ were obtained. The results showed that $\varepsilon_b$ monotonically decreases with increasing $f$. $\sigma_b$ and $W_b$ also decline with increasing $f$ when $\varphi_s$ is large. In contrast, they steeply decrease with decreasing $\varphi_s$ for small $f$ cases, and thus, they exhibit non-monotonic $f$-dependence at small $\varphi_s$ values. Compared to the case of star polymer networks, including primary loops induced an increase in $\varepsilon_b$ and decreases in $\sigma_b$ and $W_b$. However, the differences are insignificant, and the fracture behavior against $f$ and $\sigma_b$ is similar to that reported for star polymer networks. For quantitative evaluation of the difference between end-linking and star networks, the fracture characteristics were further analyzed regarding the $\xi$-dependence. The results demonstrate that $\varepsilon_b(\xi)$ collected for various $f$ and $\varphi_s$ follows the same master curve reported for star networks. $\sigma_b(\xi)$ and $W_b(\xi)$ also lie on the same master cures as those for star analogs, if they are normalized by the number density of branch points $\nu_{br}$. Universality was also observed for the broken bond fraction normalized by the branch point density. These results imply that fracture of end-linking networks is essentially the same as that for star networks and the effects of $f$ and $\varphi_s$ including loop formations are embedded in $\xi$.

The master curves for fracture characteristics against $\xi$ may propose a new aspect for network characterization. However, the examined systems and the stretching scheme are significantly different from real polymer cases; for instance, interactions including excluded volumes and osmotic forces are neglected, molecular weight distribution is not considered, and Brownian motion is omitted. The effects of these missing factors should be carefully discussed. Meanwhile, the examined situations in this study may be similar to what is observed for macroscopic elastic networks under stretch [53–55] and even for other loaded networks in general. Further studies in such directions are ongoing, and the results will be reported elsewhere.

## ACKNOWLEDGEMENTS
This study is partly supported by JST-CREST (JPMJCR1992) and JSPS KAKENHI (22H01189).